\documentclass[]{spie}  %>>> use for US letter paper
\usepackage{amsmath,amsfonts,amssymb}
\usepackage{graphicx}
\usepackage[colorlinks=true, allcolors=blue]{hyperref}
\usepackage[separate-uncertainty = true,multi-part-units=single]{siunitx} % for units \SI{100}{\micro\meter}
\usepackage[euler]{textgreek}
\usepackage{multirow}
\newcommand*{\rom}[1]{\uppercase\expandafter{\romannumeral #1\relax}}
\newcommand{\Ro}{\mathrm{R_\odot}}
\newcommand{\ud}{\,\mathrm{d}}

\title{Simulations of stray light from the surface scattering of the {S}olar {C}orona {I}mager primary mirror}

\author[a,b]{Jianchao Xue}
\author[c]{Marco Romoli}
\author[d]{Federico Landini}
\author[c]{Cristian Baccani}
\author[a,b]{Hui Li}
\author[e]{Yunqi Wang}
\author[e]{Bo Chen}
\affil[a]{Key Laboratory of Dark Matter and Space Astronomy, Purple Mountain Observatory, Chinese Academy of Sciences, 10 Yuanhua Road, Nanjing 210023, China}
\affil[b]{School of Astronomy and Space Science, University of Science and Technology of China, 96 Jinzhai Road, Hefei 230026, China}
\affil[c]{Department of Physics and Astronomy, University of Florence, Via Sansone 1, Florence 50019, Italy}
\affil[d]{Astrophysical Observatory of Torino, National Institute for Astrophysics, Via Osservatorio 20, Pino Torinese 10025, Italy}
\affil[e]{Changchun Institute of Optics, Fine Mechanics and Physics, Chinese Academy of Sciences, 3888 Dong Nanhu Road, Changchun 130033, China}

\authorinfo{Further author information: (Send correspondence to Hui Li)\\Jianchao Xue: E-mail: xuejc@pmo.ac.cn\\  Hui Li: E-mail: nj.lihui@pmo.ac.cn}

\begin{document} 
\maketitle

\begin{abstract}
The Solar Corona Imager is an internally occulted coronagraph on board the ASO-S mission, which has the advantage of imaging the inner corona in H~\rom{1} {Lyman-\textalpha} (Ly\textalpha) and white-light (WL) wavebands. However, scattering of solar disk light by the primary mirror (M1) becomes the main source of stray light. To study the methods of stray light suppression, three scattering models are used to model M1 scattering in Zemax OpticStudio. The ratio of coronal emission to predicted stray light decrease along field of view in both channels. The stray light in Ly\textalpha channel is generally lower than coronal emission, but the stray light in WL channel tends to be one order of magnitude higher than coronal signal at $2.5\,\mathrm{R_\odot}$. Optimized parameter combinations that suppress the stray light to required level are obtained, which put some limitations on the M1 manufacture. Besides, K-correlation model is recommended to simulate surface scattering.
\end{abstract}

\section{INTRODUCTION}
\label{sec:intro}  % \label{} allows reference to this section

Stray light suppression is extremely important for solar coronagraphs, because the corona is much fainter compared with the solar disk. Externally occulted coronagraphs can block solar disk light before optical systems and ensure low stray light level. They are widely used in space, such as the {Solar and Heliospheric Observatory (SOHO)/Large Angle Spectroscopic Coronagraph (LASCO)-C2, C3\cite{Bruechner1995lasco}}, the Solar Terrestrial Relation Observatory (STEREO)/Sun Earth Connection Coronal and Heliospheric Investigation (SECCHI)-COR2\cite{Howard2008secchi} and the Solar Orbiter/Multi Element Telescope for Imaging and Spectroscopy (METIS\cite{2020A&A...642A..10A}). A disadvantage of externally occulted coronagraphs is that the lower limit of the field-of-view (FoV) is usually larger than 1.5 solar radii ($\Ro$). Internally occulted coronagraphs (SOHO/LASCO-C1 and STEREO/SECCHI-COR1) can observe the inner corona down to \SI{1.1}{\Ro}, but the primary objects scattering of solar disk emission becomes the main source of stray light and prevents from observing beyond \SI{2}{\Ro}. H~\rom{1} {Lyman-\textalpha} ({Ly\textalpha}) and He~\rom{2} {Ly\textalpha} observations start to play a significant role in diagnosing solar wind and coronal mass ejections (CMEs)\cite{Romoli2017metis,Fineschi1994stray,Landini2006score,Vourlidas2019lst,Ying2019cme}, in which wavebands the brightness ratio of corona to solar disk (about \num{1e-3} for H~\rom{1} {Ly\textalpha} at \SI{1.2}{\Ro}\cite{Romoli2017stray,Verroi2012stray}) is much higher than that in white-light (WL) waveband (about \num{1e-6} at \SI{1.2}{\Ro}\cite{Romoli2017stray,Verroi2012stray}). WL observations are still important to study the K and F corona.

The \emph{Solar Corona Imager} (SCI) is one of instruments of the \emph{Lyman-\textalpha ~Solar Telescope} (LST)\cite{Li2019lst,Chen2019lst,Feng2019lst} on board the \emph{Advanced Space-based Solar Observatory} (ASO-S)\cite{Gan2019asos} mission. The SCI can image the solar corona from \SIrange[range-units=single]{1.1}{2.5}{\Ro} in both H~\rom{1} {Ly\textalpha} (\SI{121.6(100)}{\nano\meter}) and WL (\SI{700(40)}{\nano\meter}) wavebands. The SCI is an internally occulted reflecting coronagraph, and the main source of stray light is the scatttering of the solar disk light by the primary mirror (M1). To image the corona clearly, stray light is required to be fainter than coronal brightness. 

For an optical surface that fabricated by conventional techniques, its surface power spectral density (PSD) usually follows an inverse power law at high spatial frequencies\cite{Harvey2012tis}. The Lorentzian and ABg models have been used to model the distribution of light scattered by such kind of optics in coronagraphs\cite{Fineschi1994stray, Landini2006score, Sandri2018stray}. However, the applications of the two models are confined to their particular log-log slopes of the bidirectional reflectance distribution functions (BRDFs). The K-correlation model\cite{Stover1995scatter,Dittman2006abc,Harvey2012tis} can be used for an arbitrary log-log slope and is widely used for scattering analyses, but is rarely mentioned in the field of telescopes. We will compare the three scattering models after illustrating the relationship between surface scattering and surface properties.

The main purpose of this work is to predict, understand and select theoretical design solutions to suppress the SCI stray light that due to the M1 surface scattering. The SCI optical design is described in Section~\ref{sect:inst}. The methods of Stray light simulations, including theories of surface scattering and scattering models, are introduced in Section~\ref{sect:meth}. Simulation results and analyses are shown in Section~\ref{sect:result}, followed by conclusions and discussion in Section~\ref{sect:conc}.

%%%%%%%%%%%%%%%%%%%% Instrument description %%%%%%%%%%%%%%%%%%%%%%%%%
\section{Instrument description}
\label{sect:inst}
\begin{figure}[htbp]
\begin{center}
\includegraphics[width=5in]{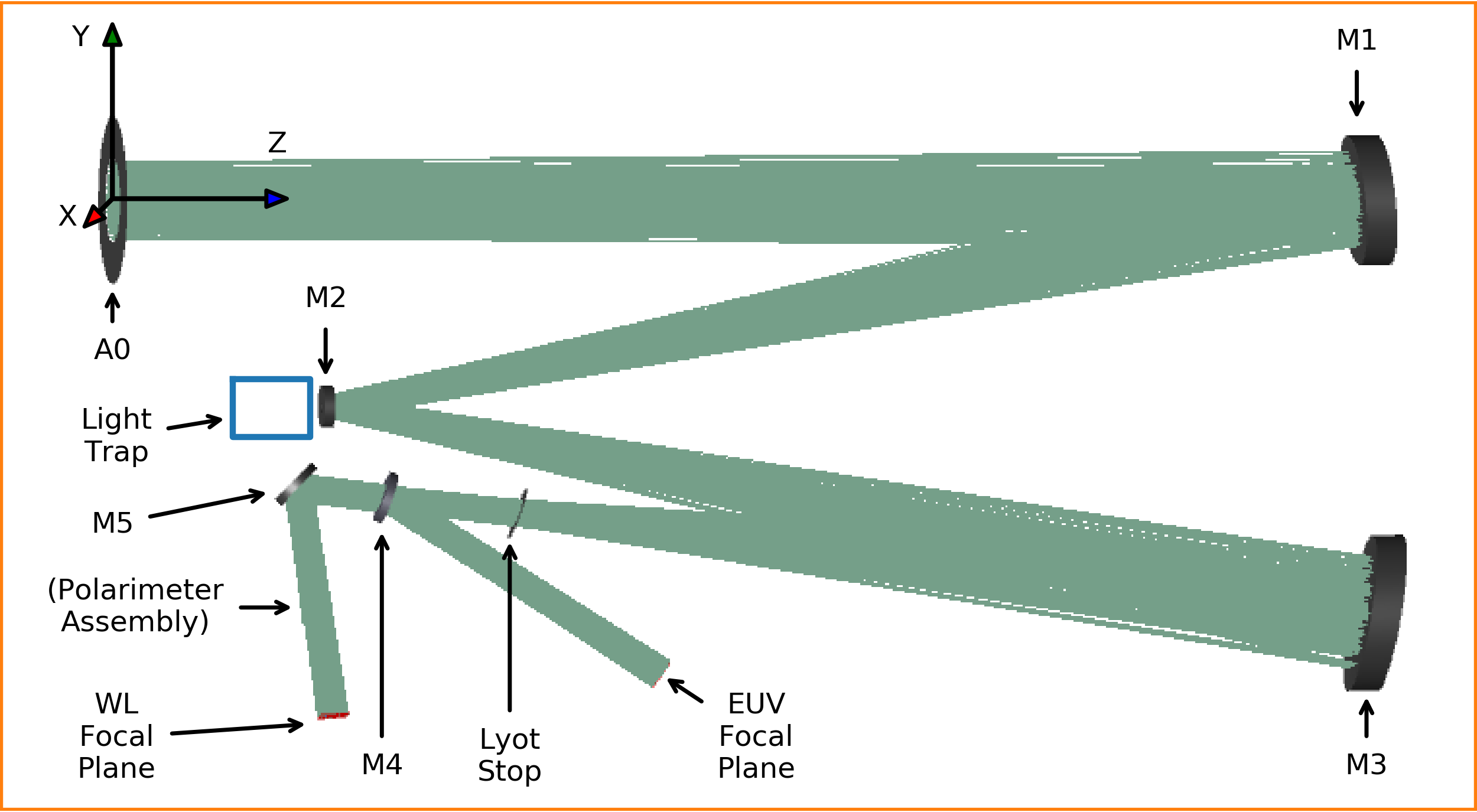}
\end{center}
\caption[Schematic layout of the SCI coronagraph.]
{\label{fig:sci}
Schematic optical layout of the SCI coronagraph. Polarimeter assembly in the WL channel is not shown.}
\end{figure}

The SCI is a Lyot-type reflecting and off-axis coronagraph. It has two channels to image corona in WL and H~\rom{1} {Ly\textalpha} wavebands simultaneously. The SCI optical design is shown in Fig.~\ref{fig:sci}. The SCI has an entrance aperture (A0) with diameter of \SI{60}{\mm}, and realizes an effective focal length of \SI{945}{\mm} for both channels. The optical axis of the beam path, from A0 to the final focal planes, lays on a plane which will be called \emph{optical axis plane} (if no additional instruction, all the following ``focal plane(s)'' refer to the ``final focal plane(s)''). Both solar coronal and disk radiation enter the system through A0 and are imaged by the primary mirror (M1) onto the secondary mirror (M2) plane. Solar disk light enters the cone-shaped hole in the center of M2, as the inner occulter, and is absorbed by the light trap. The beam splitter (M4), coated with magnesium fluoride, is used to reflect {Ly\textalpha} emission and transmit WL to their detectors. A Lyot stop (LS) is placed on the conjugate plane of A0 imaged by M1, M2 and the 3rd mirror (M3), to block diffracted light from the A0 edge. The polarimeter assembly is arranged between the plane mirror M5 and the WL detector, but not shown in Fig.~\ref{fig:sci}. Both M2 and square detectors function as the field stops and enable the SCI FoV to be \SIrange[range-units=single, range-phrase = --]{1.1}{2.5}{\Ro} at \SI{1}{AU}.

The \emph{global reference system} (GRS) used in this work is annotated in Fig.~\ref{fig:sci}. The origin of the axes is at the center of A0; Z-axis is along the optical axis from A0 towards M1; Y-axis is directed away from the off-axis direction on the optical axis plane; X-axis completes the GRS.

%%%%%%%%%%% Method %%%%%%%%%%%%%%%%%
\section{Stray light simulations and surface scattering models}
\label{sect:meth}
We conduct stray light simulations in the non-sequential modality of Zemax OpticStudio. The {Ly\textalpha} and WL channels are simulated separately with the wavelengths of \SI{122}{\nm} and \SI{700}{\nm}, respectively. M4 is treated as a mirror in the {Ly\textalpha} channel and as a lens in the WL channel with the material of silica. Only the rays scattered by M1 and reflected by M2 are traced to improve the simulation efficiency.

Since scattering of primary mirror M1 contributes the majority of the stray light, we focus on M1's scattering in our simulations and ignore other sources of stray light, such as the diffracted light from the A0 edge and subsequently scattered by optics, the solar disk radiation diffracted and/or scattered by the light trap, scattering of mechanical surfaces, and contamination. We have conducted both the simulations with and without including M2 scattering and found that their results are almost the same. In this paper, we only show the results without considering other optics' scattering except for M1 surface for simplicity. 

Some basic knowledge about surface scattering is necessary to conduct the simulations and understand the results. Here we briefly introduce the scattering theories and models associated with this work. The definitions of scattering quantities used in this work are summarized in Appendix~\ref{sect:defi}.

%%%% PSD and BRDF %%%%
\subsection{Surface properties and scattering distributions} 
\label{subs:scat}

\begin{figure}[htbp]
\begin{center}
\includegraphics[width=4in]{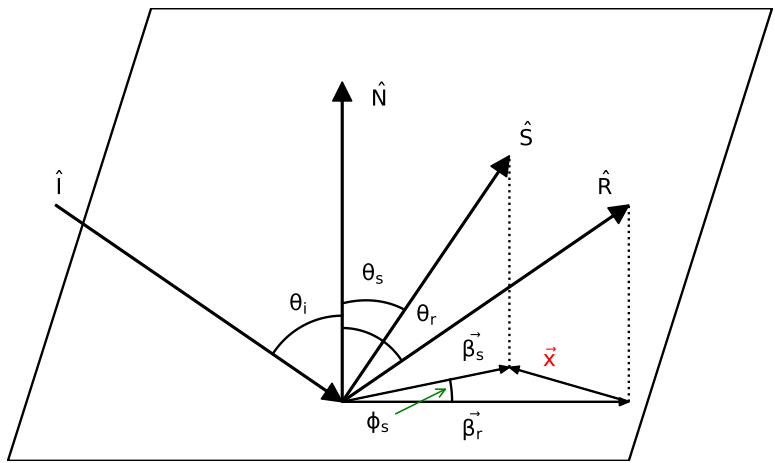}
\end{center}
\caption[Interpretation of x vector.]
{\label{fig:xvec}
Interpretation of $\vec{x}$. $\hat{N}$, $\hat{I}$, $\hat{R}$, $\hat{S}$ are unit vectors of surface normal and incident, specular reflected, scattered rays. $\vec{\beta_r}$, $\vec{\beta_s}$ are projections of $\hat{R}$, $\hat{S}$ on the scattering surface.
}
\end{figure}

Surface \emph{power spectral density} (PSD) function and \emph{bidirectional scattering distribution function} (BSDF), including \emph{bidirectional reflectance distribution function} (BRDF) for reflective samples, are usually used to describe surface properties and scattering distributions, respectively. For ``an arbitrary, smooth, clean, front-surface reflector'', the surface PSD function and BRDF can be related by the Rayleigh-Rice perturbation theory (P.~85 in Ref.~\citenum{Stover1995scatter}) and grating equation:
\begin{equation}
\label{eq:rayl}
\mathrm{BRDF}(\lvert\vec{x}\rvert)=\frac{16\pi^2}{\lambda^4}\cos\theta_i\cos\theta_sQS_2(f)\, ,
\end{equation}
and
\begin{equation}
\label{eq:grat}
\lvert\vec{x}\rvert=f\lambda \, .
\end{equation}
In Eqs.~(\ref{eq:rayl}) and~(\ref{eq:grat}), $\lambda$ is light wavelength, $Q$ is reflectivity polarization factor, $S_2(f)$ is two-dimensional (2-D) PSD function along spatial frequency $f$, $\theta$ is polar angle relative to the scattering surface normal, and subscripts $i$, $s$, $r$ are used, or will be used, to represent quantities of incidence, scattering and specular reflection, respectively. $\vec{x}$ is the difference between projections of scattered and specular reflected ray vectors down to the scattering surface, as shown in Fig.~\ref{fig:xvec}. In Fig.~\ref{fig:xvec}, $\vec{\beta_r}$ and $\vec{\beta_s}$ are projections of $\hat{R}$ and $\hat{S}$, and defined by $\lvert\vec{\beta_r}\rvert=\sin\theta_r=\sin\theta_i$ and $\lvert\vec{\beta_s}\rvert=\sin\theta_s$, respectively. $\vec{x}$ is defined by 
\begin{equation} \label{eq:xvec}
\vec{x}=\vec{\beta_s}-\vec{\beta_r} \,.
\end{equation} 
The well-known 1-D grating equation is 
\begin{equation} \label{eq:1dgrat}
\sin\theta_n = \sin\theta_i + n f \lambda \, , 
\end{equation}
where $n$ is an integer. Analogously, the hemispherical grating equation at the first-order diffraction position is expressed as (P.~75 in Ref.~\citenum{Stover1995scatter})

\begin{equation} \label{eq:2dgrat}
\vec{x} = \binom{\sin\theta_s \cos\phi_s - \sin\theta_i}{\sin\theta_s\sin\phi_s} = \binom{f_x}{f_y}\lambda = \vec{f}\lambda \, ,
\end{equation}
where $f_x$ is the surface spatial frequency on the plane of incidence, and $f_y$ is along the direction perpendicular to the plane of incidence. By Eq.~(\ref{eq:2dgrat}), Eq.~(\ref{eq:grat}) is obtained. Here the surface roughness is assumed to be isotropic and $S_2(\vec{f})$ is symmetrical, but BRDF is asymmetrical along $\theta_s$ and $\phi_s$ (azimuth angle) except for normal incidence.

\begin{figure}[htbp]
\begin{center}
\includegraphics[width=6in]{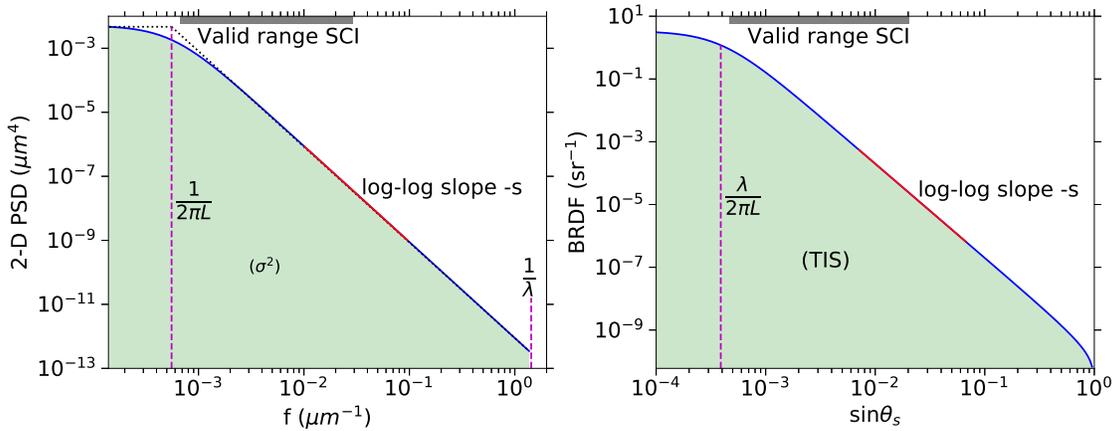}
\end{center}
\caption[Examples of surface PSD and BRDF following inverse power laws.]
{\label{fig:psdb}
    Examples of (a) a PSD function and (b) a BRDF. The data are from K-correlation function with $\sigma = 0.1\,\rm{nm}$, $B=$\SI{1800}{\micro\meter}, $s=3$ and normal incidence.
}
\end{figure}

Optical surfaces fabricated by conventional polishing techniques tend to have the PSD with a plateau at low spatial frequencies and following an inverse power law at high frequencies\cite{Harvey2012tis}. A PSD function and BRDF of \emph{K-correlation} model\cite{Stover1995scatter,Dittman2006abc,Harvey2012tis} are shown in Fig.~\ref{fig:psdb} as an example. The profiles are characterized by their integrals, breakpoints and log-log slopes.

\begin{itemize}
\item Integrals. The integral of the PSD with limit of $f=1/\lambda$ is the square of \emph{relevant band-limited RMS roughness} $\sigma$\cite{Harvey2012tis} (or \emph{total effective RMS roughness} in Ref.~\citenum{Dittman2006abc}):

\begin{equation}
\label{eq:sigma}
\sigma^2(\lambda) = \iint_{-1/\lambda,-1/\lambda}^{1/\lambda,1/\lambda} S_2(f_x,f_y)\ud f_x\ud f_y = 2\pi\int_0^{1/\lambda}S_2(f)f\ud f \, .
\end{equation}
The right part of Eq.~(\ref{eq:sigma}) is applicable for isotropic surface roughness and normal incidence. The relevant band-limited RMS roughness is used to evaluate effective surface roughness for a certain light with wavelength $\lambda$. The upper limit in Eq.~(\ref{eq:sigma}) is due to that the spatial frequencies larger than $1/\lambda$ correspond to $\theta_s >90\,^\circ$ for normal incidence, by Eq.~(\ref{eq:1dgrat}), and don't contribute to optical scattering. It means that the relevant band-limited RMS roughness for a same surface used in different wavelength channels, {Ly\textalpha} and WL for SCI, are different. Therefore, in our simulations once $\sigma$ is given in WL channel, the corresponding roughness in {Ly\textalpha} channel is necessary to be derived by \emph{wavelength scaling laws}\cite{Dittman2006abc}. 

The \emph{total integrated scatter} (TIS) equals to the integral of BRDF along solid angle with the correction factor $\cos\theta_s$:

\begin{equation}
\label{eq:tis}
\mathrm{TIS} = \iint_{\Omega_s} \mathrm{BRDF} \cos\theta_s \ud\Omega_s = \iint_{\Omega_s} \mathrm{BRDF} \cos\theta_s \sin\theta_s \ud\varphi_s\ud\theta_s \, .
\end{equation}
TIS characterizes the total amount of scattered light in the scattering hemisphere. By substituting Eqs.~(\ref{eq:rayl}),~(\ref{eq:grat}),~(\ref{eq:sigma}) into~(\ref{eq:tis}) and assuming $\cos\theta_i\cos\theta_sQ=1$, we obtain the approximate TIS:
\begin{equation}
\label{eq:tis_sigma}
\mathrm{TIS} \approx \left( \frac{4\pi\sigma}{\lambda}\right)^2\, .
\end{equation}
 Equation~(\ref{eq:tis_sigma}) indicates that the total scattering is mainly determined by the square of $\sigma$ for a certain wavelength. Notice that Eq.~(\ref{eq:tis_sigma}) is derived without considering the PSD profile, but only on the basis of the Rayleigh-Rice perturbation theory Eq.~(\ref{eq:rayl}).

\item Breakpoints (P.105 in Ref.~\citenum{Stover1995scatter}) or roll-off frequency/{angle\cite{Sandri2018stray}}. As defined in Fig.~\ref{fig:psdb}(a), roll-off frequency can be considered the amplitude of the circular roughness distribution; roughness almost keeps a constant within the circle and drops following an inverse power law beyond it. Analogously, roll-off angle can be considered the amplitude of the conical scattering distribution. Roll-off frequency and angle are expressed as:

\begin{equation}
\label{eq:rollf}
f_0 = \frac{1}{2\pi L}\, ,
\end{equation}
and
\begin{equation}
\label{eq:rolla}
\sin\theta_0 = \frac{\lambda}{2\pi L}\, .
\end{equation}

$L$ in Eqs.~(\ref{eq:rollf}) and~(\ref{eq:rolla}) is related to the \emph{autocorrelation length} $l_c$, but the relationship depends on the value of log-log slope -s. Dittman\cite{Dittman2006abc} derived the relationship by Fourier transform, and found that $L \sim 5l_c$ (the coefficient is \num{5.3+-0.2}) at $s=2$ and $L=l_c$ at $s=3$. Following Dittman\cite{Dittman2006abc}, $L$ is called the \emph{typical surface wavelength}.

\item Log-log slopes -s. The log-log slope determines the dropping speed of PSD (BRDF) at high spatial frequencies (large scattering angles). The deviation of the BRDF at large scattering angles from the inverse power law is due to the factor $\cos\theta_s$ in Eq.~(\ref{eq:rayl}).
\end{itemize}

The stray light we studied in this work shares the same optical path after M1 as the coronal signal; otherwise, the scattered light cannot be reflected by M2 or be received by detectors. Therefore, the effective range of scattering angles on M1 surface is limited, and can be derived from the FoV of the SCI. The SCI FoV at \SI{1}{AU} is about \SIrange[range-units=single, range-phrase = --]{1.1}{2.5}{\Ro}. If the square detector corners are considered, the FoV reaches about \SI{3.36}{\Ro}. The angular radius of the sun is about $0.266\,^\circ$. Considering oblique incidence, the lower limit of the effective scattering angle is $(1.10-1.00)\,\Ro \times 0.266\,^\circ/\Ro \approx 0.027\,^\circ$ and the upper limit is $(3.36+1.00)\,\Ro \times 0.266\,^\circ/R_\odot \approx 1.2\,^\circ$. By Eq.~(\ref{eq:1dgrat}) for the first-order diffraction and assuming normal incidence, the corresponding effective spatial frequencies of M1 surface in the WL channel are \SIrange[range-units=single, range-phrase = --]{6.7e-4}{2.9e-2}{\per\um}, and the spatial wavelengths are \SIrange[range-units=single, range-phrase = --]{1.5}{0.034}{\mm}.

%%%%%%%%%%%%%%%% Scattering models %%%%%%%%%%%%
\subsection{Surface scattering models}
\label{subs:model}
Following Refs.~\citenum{Fineschi1994stray, Landini2006score, Sandri2018stray, Harvey2012tis, Stover1995scatter, Dittman2006abc}, we assume that the BRDF of SCI M1 surface has a plateau at small scattering angles and follows an inverse power law at large scattering angles, as described in Section~\ref{subs:scat}. There are three models that were used to simulate such kind of surface scattering. They are K-correlation, ABg and Lorentzian models. Since $\sigma$ of M1 surface in {Ly\textalpha} and WL channels are different, both BRDF and wavelength scaling laws are needful in our simulations.

\begin{enumerate}
%% K-correlation %%
 \item K-correlation (or ABC) {model\cite{Dittman2006abc}}. BRDF:
  \begin{subequations} \label{eq:kbrdf}
   \begin{align}
   \mathrm{BRDF}(\lvert\vec{x}\rvert, s \neq 2)&=\frac{2\pi\Delta n^2RB^2}{\lambda^4}\cdot\frac{\sigma^2(s-2)}{1-\left[1+B^2/\lambda^2\right]^{1-s/2}}\cdot\frac{\cos\theta_i\cos\theta_s}{\left[1+\left(B\lvert\vec{x}\rvert/\lambda\right)^2\right]^{s/2}} \, , \label{eq:kbsn2}\\
   \mathrm{BRDF}(\lvert\vec{x}\rvert, s = 2)&=\frac{4\pi\Delta n^2RB^2}{\lambda^4}\cdot\frac{\sigma^2}{\ln\left(1+B^2/\lambda^2\right)}\cdot\frac{\cos\theta_i\cos\theta_s}{1+\left(B\lvert\vec{x}\rvert/\lambda\right)^2} \, . \label{eq:kbse2}
    \end{align}
   \end{subequations}
   Scaling laws:
  \begin{subequations}\label{eq:kscal}
   \begin{align}
   \textrm{For $s \neq 2$,}\quad \frac{\sigma^2(\lambda_2)}{\sigma^2(\lambda_1)}&=\frac{1-\left[1+\left(B/\lambda_2\right)^2\right]^{1-s/2}}{1-\left[1+\left(B/\lambda_1\right)^2\right]^{1-s/2}} \, ,\label{eq:kssn2}\\
   \textrm{For $s = 2$,}\quad \frac{\sigma^2(\lambda_2)}{\sigma^2(\lambda_1)}&=\frac{\ln\left[1+\left(B/\lambda_2\right)^2\right]}{\ln\left[1+\left(B/\lambda_1\right)^2\right]} \, .\label{eq:ksse2}
   \end{align}
  \end{subequations}

In Eqs.~(\ref{eq:kbrdf}) and~(\ref{eq:kscal}), $B=2\pi L$ is the inverse of roll-off frequency, $\rm{\Delta} n=2$ is the refractive index change, and $R=1$ is the surface reflectivity. Additionally, roll-off angle can be derived from Eq.~(\ref{eq:kbrdf}) by $\left(B\lvert\vec{x}\rvert/\lambda\right)^2=1$.

%% ABg %%
 \item ABg {model\cite{Dittman2002cont,Sandri2018stray}}. BRDF:
  \begin{equation}
   \label{eq:abrdf}
    \mathrm{BRDF}(\lvert\vec{x}\rvert)=\frac{\mathrm{\Delta} n^2}{8\pi} \cdot \frac{\left(\frac{2\pi}{\lambda}\right)^4\sigma^2L^2}{1+\left(\frac{2\pi}{\lambda}L\lvert\vec{x}\rvert\right)^2}=\frac{A}{B_a+\lvert\vec{x}\rvert^g} \, ,
   \end{equation}
   with
   \begin{equation}
    \label{eq:abg}
     A=\frac{\pi}{2}\cdot\frac{\mathrm{\Delta} n^2\sigma^2}{\lambda^2},\quad B_a=\left(\frac{\lambda}{2\pi L}\right)^2,\quad g=2\, .
    \end{equation}
   Scaling laws (P.251 in Ref.~\citenum{Zemax2011manual}):
    \begin{equation} \label{eq:ascal}
     \frac{A(\lambda_2)}{A(\lambda_1)} = \left(\frac{\lambda_2}{\lambda_1}\right)^{g-4}, \quad \frac{B_a(\lambda_2)}{B_a(\lambda_1)} = \left(\frac{\lambda_2}{\lambda_1}\right)^{g} \, .
    \end{equation}
Equation~(\ref{eq:ascal}) indicates that $A$ and $B_a$ expressions depend on the log-log slope $-g$. The expressions in Eq.~(\ref{eq:abg}) are only applicable to $g=2$.
%% Lorentzian %%
 \item Lorentzian {model\cite{Fineschi1994stray,Landini2006score}}. BRDF:
  \begin{equation} \label{eq:lbrdf}
   \mathrm{BRDF}(\lvert\vec{x}\rvert) = \left(\frac{4\pi\sigma}{\lambda}\right)^2\cdot\frac{1}{2\pi}\cdot\frac{(2\pi l_c)^2}{\left[1+(2\pi l_c/\lambda)^2\lvert\vec{x}\rvert^2\right]^{3/2}\cos\theta_s} \,.
  \end{equation}
  Scaling law:
  \begin{equation} \label{eq:lscal}
   \frac{\sigma^2(\lambda_2)}{\sigma^2(\lambda_1)}=\frac{1-\left[1+\left(B/\lambda_2\right)^2\right]^{-1/2}}{1-\left[1+\left(B/\lambda_1\right)^2\right]^{-1/2}} \, .
  \end{equation}
 We derived Eq.~(\ref{eq:lscal}) by substituting Eq.~(9) in Ref.~\citenum{Fineschi1994stray} into Eq.~(\ref{eq:sigma}). Because Eq.~(\ref{eq:lbrdf}) is a Lorentzian function, here this model is named by "Lorentzian model".
\end{enumerate}

\begin{figure}[htbp]
\begin{center}
\includegraphics[width=6in]{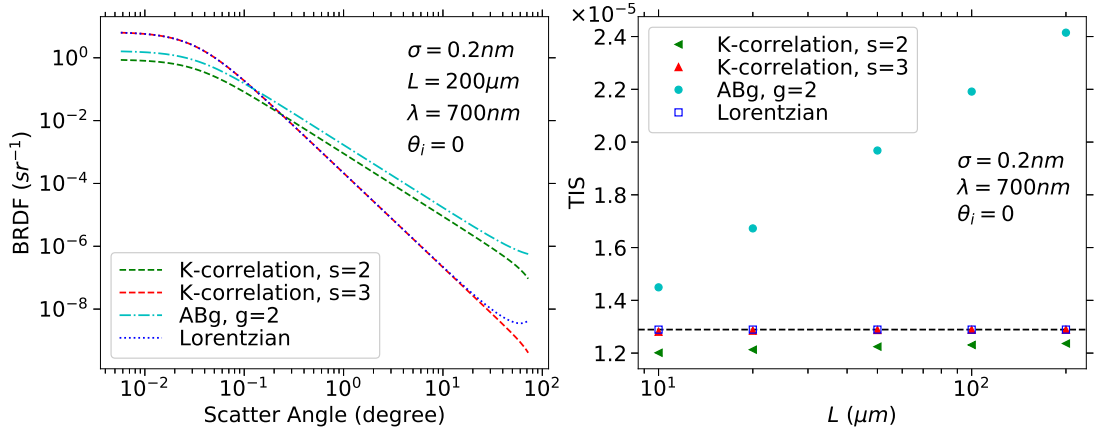}
\end{center}
\caption[BRDF and TIS of three scattering models.]
{\label{fig:brdft}
    (a) BRDFs of the three scattering models. (b) TISs calculated by integrating BRDFs along solid angle (Eq.~\ref{eq:tis}, symbols) for different scattering models, and by the approximation function (Eq.~\ref{eq:tis_sigma}, dashed line). Model parameters are noted.}
\end{figure}

To compare the three scattering models, their BRDFs and TISs are plotted in Fig.~\ref{fig:brdft}. In Fig.~\ref{fig:brdft}(a), BRDFs are plotted in log-log scale. The K-correlation model has the same profile as the ABg model for slope of \num{-2}, and changes to the Lorentzian model when slope is \num{-3}\cite{Stover1995scatter,Dittman2006abc}, except for the large scattering angle regions due to that ABg and Lorentzian models exclude the correction factor $\cos \theta_s$. In Fig.~\ref{fig:brdft}(b), the large deviation of the ABg TIS (derived from Eq.~\ref{eq:tis}) from the approximate TIS (by Eq.~(\ref{eq:tis_sigma})) suggests the deviation of ABg model from the Rayleigh-Rice perturbation theory. Both the TIS of K-correlation and Lorentzian model are similar to the approximate TIS, due to that they are derived on the basis of the Rayleigh-Rice perturbation theory\cite{Dittman2006abc,Fineschi1994stray}. Since the K-correlation and ABg models have the same profile for slope of \num{-2} when their BRDFs are plotted in log-log scale, the TIS difference between the two models can be eliminated by adjusting $\sigma$ value in K-correlation model, or by adjusting $A$ value in ABg model. Therefore, the ABg (when $g=2$) and Lorentzian models can be replaced by the K-correlation model. An advantage of the K-correlation model is that an arbitrary log-log slope can be used. By comparison, the Lorentzian model has a log-log slope of \num{-3}. If $g\neq 2$ in ABg model, Eq.~(\ref{eq:abg}) fails and the relationships between $A$, $B_a$ and $\sigma$, $L$ should be redefined for different $g$ values, which is difficult and inconvenient. 

The integrals are conducted by the double exponential method\cite{Landini2006score}, and the routine code is available online (\url{http://www.kurims.kyoto-u.ac.jp/~ooura/intde.html}).

%%%%%%%%%%%%%%%% Results %%%%%%%%%%%%%%%%
\section{Results and analyses}
\label{sect:result}

%%%% Stray light level %%%%
\subsection{Predicted stray light levels with common parameters}
\label{subs:level}

\begin{table}[h]
\caption{Surface and model parameters to predict stray light levels.}
\label{tab:case}
\begin{center}
\begin{tabular}{ |c|l|l|l|l|l| }
\hline
\multirow{2}{*}{Slope} & \multirow{2}{2.8cm}{Surface Property} & \multirow{2}{2.5cm}{Model} & \multirow{2}{3cm}{Model Parameter} & \multicolumn{2}{c|}{TIS}\\
                   \cline{5-6}
                      & & & & \SI{700}{\nm}  & \SI{122}{\nm}\\
\hline
\rule[-1ex]{0pt}{3.5ex} \multirow{6}{*}{s=2} & $\sigma = 0.1$, $l_c=10$ & K-correlation & $B=314$ & \num{3.06e-6} & \num{1.31e-4} \\
                   \cline{2-6}
\rule[-1ex]{0pt}{3.5ex}                     & $\sigma = 0.2$, $l_c=10$ & K-correlation & $B=314$ & \num{1.22e-5} & \num{5.24e-4} \\
                   \cline{2-6}
\rule[-1ex]{0pt}{3.5ex}                     & $\sigma = 0.3$, $l_c=10$ & K-correlation & $B=314$ & \num{2.75e-5} & \num{1.18e-3} \\
                    \cline{2-6}
\rule[-1ex]{0pt}{3.5ex}                     & $\sigma = 0.2$, $l_c=40$ & K-correlation & $B=1.26\times10^3$ & \num{1.22e-5} & \num{5.24e-4} \\
                   \cline{2-6}
\rule[-1ex]{0pt}{3.5ex}                     & \multirow{2}{2.4cm}{$\sigma = 0.2$, $l_c=10$} & \multirow{2}{1.9cm}{ABg} & $A=5.13\times10^{-7}$ & \multirow{2}{2cm}{\num{1.97e-5}} & \multirow{2}{2cm}{\num{8.33e-4}} \\
\rule[-1ex]{0pt}{3.5ex}                    & & & $B_a=4.96\times10^{-6}$ & & \\
                   \hline
\rule[-1ex]{0pt}{3.5ex} \multirow{5}{*}{s=3} & $\sigma = 0.1$, $l_c=10$ & K-correlation & $B=62.8$ & \num{3.20e-6} & \num{1.07e-4} \\
                   \cline{2-6}
\rule[-1ex]{0pt}{3.5ex}                     & $\sigma = 0.2$, $l_c=10$ & K-correlation & $B=62.8$ & \num{1.28e-5} & \num{4.28e-4} \\
                   \cline{2-6}
\rule[-1ex]{0pt}{3.5ex}                     & $\sigma = 0.3$, $l_c=10$ & K-correlation & $B=62.8$ & \num{2.88e-5} & \num{9.63e-4} \\
                    \cline{2-6}
\rule[-1ex]{0pt}{3.5ex}                     & $\sigma = 0.2$, $l_c=40$ & K-correlation & $B=251$ & \num{1.28e-5} & \num{4.25e-4} \\
                   \cline{2-6}
\rule[-1ex]{0pt}{3.5ex}                     & $\sigma = 0.2$, $l_c=10$ & Lorentzian & - & \num{1.29e-5} & \num{4.28e-4} \\
\hline
%\specialrule{0em}{1pt}{0pt}
\multicolumn{6}{|l|}{Units: $[\sigma] -$\si{nm}; ~$[l_c]$ and $[B]-$\si{\um}; ~$[TIS]$, $[s]$, $[B_a]$, $[g] -$\num{1}; ~$[A] -$\num{1} or \si{\steradian^{-1}}.} \\
%\specialrule{0em}{1pt}{0pt}
\hline
%\specialrule{0em}{1pt}{0pt}
\multicolumn{6}{|l|}{Reference wavelength for $\sigma$, $A$ and $B_a$ values: \SI{700}{\nm}.}  \\ 
%\specialrule{0em}{1pt}{0pt}
\hline
\end{tabular}
\end{center}
\end{table}

\begin{figure}[htbp]
\begin{center}
\includegraphics[width=6in]{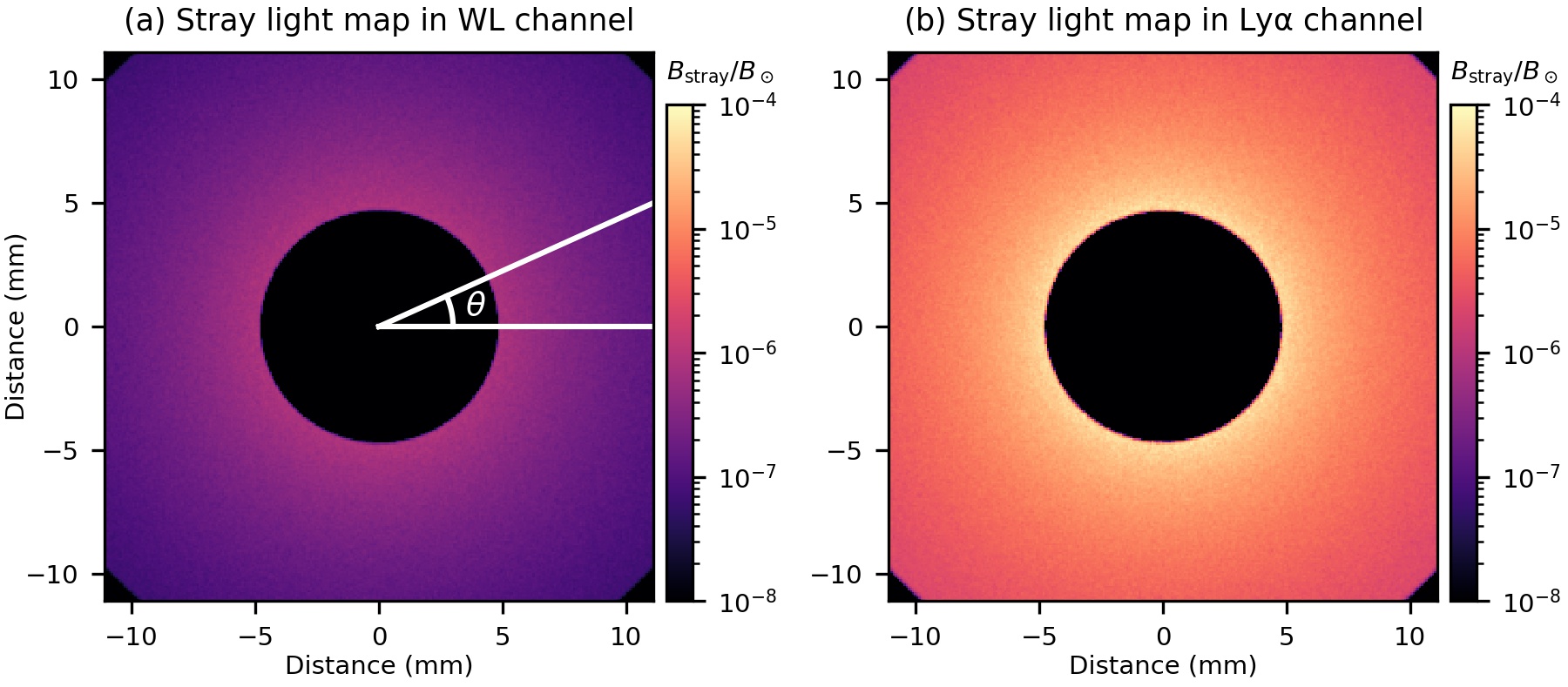}
\end{center}
\caption[Stray light maps.]
{\label{fig:slmap}
    Predicted stray light map on focal planes in WL (a) and Ly\textalpha (b) channels, respectively. K-correlation model is used with $\sigma=0.2\,\mathrm{nm}$, $l_c=10\,\mathrm{\mu m}$, and $s=3$ (second case in Table~\ref{tab:case}).
}
\end{figure}

\begin{figure}[htbp]
\begin{center}
\includegraphics[width=6in]{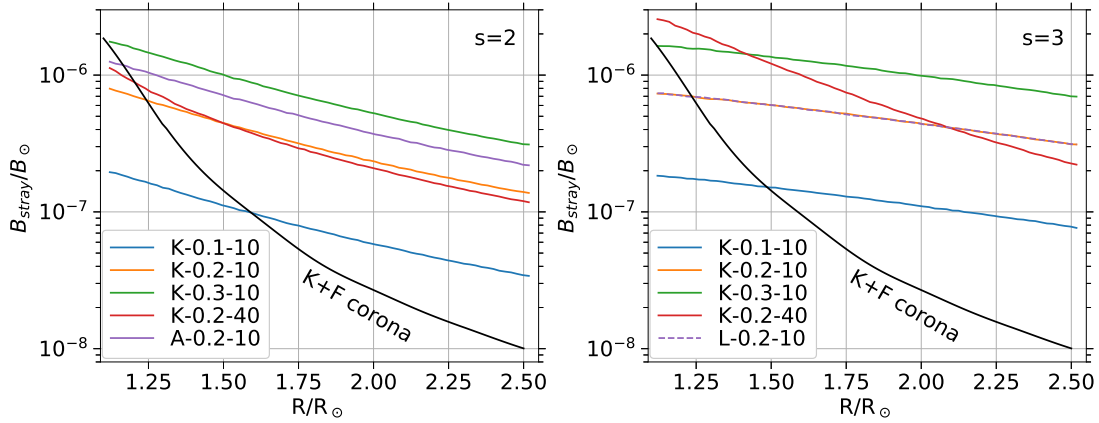}
\end{center}
\caption[Stray light predictions in WL channel.]
{\label{fig:sllwl}
    Predicted stray light of the WL channel along distance with log-log slope of \num{-2} (a) and \num{-3} (b), respectively. Labeled parameters in legends are the scattering models (``K'' for K-correlation, ``A'' for ABg, and ``L'' for Lorentzian model), $\sigma$ in units of \si{nm}, and $l_c$ in units of \si{\um}, see Table~\ref{tab:case}. K and F corona intensities are from Ref.~\citenum{Romoli2017stray}.
}
\end{figure}

\begin{figure}[htbp]
\begin{center}
\includegraphics[width=6in]{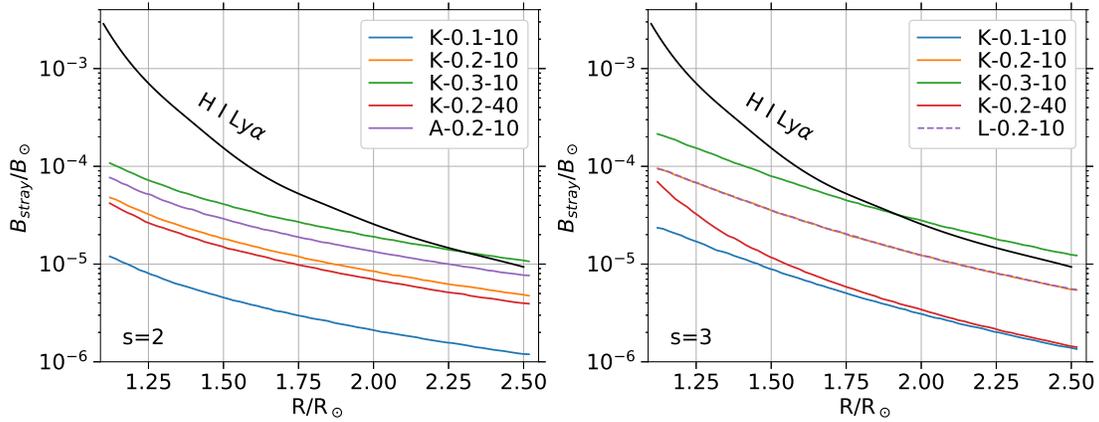}
\end{center}
\caption[Stray light predictions in {Ly\textalpha} channel.]
{\label{fig:slleuv}
    Similar to Fig.~\ref{fig:sllwl} but for {Ly\textalpha} channel. H \rom{1} {Ly\textalpha} intensities are from Ref.~\citenum{Verroi2012stray}.
}
\end{figure}

A uniform solar disk ("source two angle" in non-sequential modality of Zemax) is simulated with angular extension of \ang{0.266}, \num{1e7} rays, and placed just before the entrance aperture A0. Image distortion and ghosts (produced by the rays retroreflected by M4) are found to have little effect on the results and are ignored. Both the stray light irradiance $B_{\rm{stray}}$ and solar disk irradiance $B_\odot$ are evaluated on the focal planes. $B_\odot$ is obtained with ideal surfaces and the hole in M2 removed, and $B_{\rm{stray}}$ is evaluated by applying a micro-roughness on the primary mirror only. The assumed parameters that we adopt to define surface roughness have been commonly measured in laboratory or used by others: $\sigma$ of \SIrange[range-units=single, range-phrase = --]{0.1}{0.3}{\nm}\cite{Bruechner1995lasco,Sandri2018stray}, $l_c$ of \SIrange[range-units=single, range-phrase = --]{10}{40}{\um}\cite{Fineschi1994stray,Dittman2006abc,Sandri2018stray}, $s$ of \num{2}\cite{Dittman2002cont,Sandri2018stray} and \num{3}\cite{Fineschi1994stray,Landini2006score}. K-correlation scattering model is mainly used, and ABg and Lorentzian models are used to check the simulation results. Cases with different surface properties and scattering models are listed in Table~\ref{tab:case}. 

An example of stray light distributions on focal planes in both WL and Ly\textalpha channels are shown in Fig.~\ref{fig:slmap}. They are the results of the second case in Table~\ref{tab:case}. Both the ratios of  $B_{\rm{stray}}$  to $B_\odot$ in the two channels decrease along FoV, and $B_{\rm{stray}}/B_\odot$ is larger in Ly\textalpha channel than that in WL channel. The obtained 2-D stray light distributions are averaged along central angle $\theta$ (Fig.~\ref{fig:slmap}) to obtain stray light curves along distance $R$. $B_{\rm{stray}}$ is expressed in units of $B_\odot$ and $R$ is in units of $\Ro$ at \SI{1}{AU}. The results are plotted in Fig.~\ref{fig:sllwl} for WL channel and in Fig.~\ref{fig:slleuv} for {Ly\textalpha} channel.

The predicted stray light in WL channel tends to be brighter than the K + F corona. Especially at far FoV, the stray light is around one order of magnitude higher than coronal emission at $2.5\,\Ro$. In {Ly\textalpha} channel, it's relatively easier to fulfill the stray light requirements due to the lower contrast ratio between disk and coronal emission, though the stray light is higher than that in the WL channel for the same incident intensity. In both channels, the signal-to-noise ratios (S/Ns) decrease obviously along FoV.

As generally known, reducing roughness can efficiently suppress the stray light. The effect of $l_c$ and $s$ on stray light distributions are a little more complicate. When $\sigma$ is a constant, larger $l_c$ (smaller roll-off frequency) or larger $s$ (smaller slope) lead to a lower roughness at high spatial frequencies, but a higher roughness at low frequencies, since the integral of the surface PSD function equals to $\sigma$. In addition to surface properties, stray light distributions also depend on the incident waveband, because $\lambda=122\,\rm nm$ corresponds to a smaller roll-off angle than $\lambda=700\,\rm nm$ from Eq.~(\ref{eq:rolla}). Although larger $s$ results in higher stray light in both the channels, stray light distributions get gentler in WL channel, and get steeper in {Ly\textalpha} channel.

The simulated stray light using ABg model is higher than that using K-correlation model, but they have comparable profiles. The stray light simulated by Lorentzian and K-correlation model are completely the same. These results are in agreement with the comparison among the scattering models made in Section~\ref{subs:model}.

%%%% Fitting %%%%%
\subsection{Optimized parameter combinations}
\label{subs:fit}

\begin{table}[h]
\caption{Optimized surface parameters of SCI primary mirror that can meet stray light requirements.}
\label{tab:slfit}
\begin{center}
\begin{tabular}{|c|c|c|c|}
\hline
\rule[-1ex]{0pt}{3.5ex} \multirow{2}{*}{Model Parameter} & \multirow{2}{3cm}{Autocorrelation Length} & \multicolumn{2}{c|}{TIS}\\
                   \cline{3-4}
                      & & \SI{700}{\nm}  & \SI{122}{\nm}\\
\hline
\rule[-1ex]{0pt}{3.5ex} $\sigma=0.1$, $B=1800$, $s=3$ & $l_c=286\pm0.5$ & \multirow{2}{*}{\num{3.22e-6}} & \multirow{2}{*}{\num{1.06e-4}} \\
\cline{1-2}
\rule[-1ex]{0pt}{3.5ex} $\sigma=0.1$, $B=1300$, $s=3.2$ & $l_c=228\pm33$ &  & \\
\hline
\rule[-1ex]{0pt}{3.5ex} $\sigma=0.12$, $B=1800$, $s=3.2$ & $l_c=315\pm45$ & \num{4.64e-6} & \num{1.53e-4}\\
\hline
\multicolumn{4}{|l|}{Units: $[\sigma] -$\si{nm}; ~$[l_c]$ and $[B]-$\si{\um}; ~$[TIS]$, $[s] -$\num{1}.} \\
\hline
\multicolumn{4}{|l|}{Reference wavelength for $\sigma$ values: \SI{700}{\nm}.}  \\ \hline
\end{tabular}
\end{center}
\end{table}

\begin{figure}[htbp]
\begin{center}
\includegraphics[width=6in]{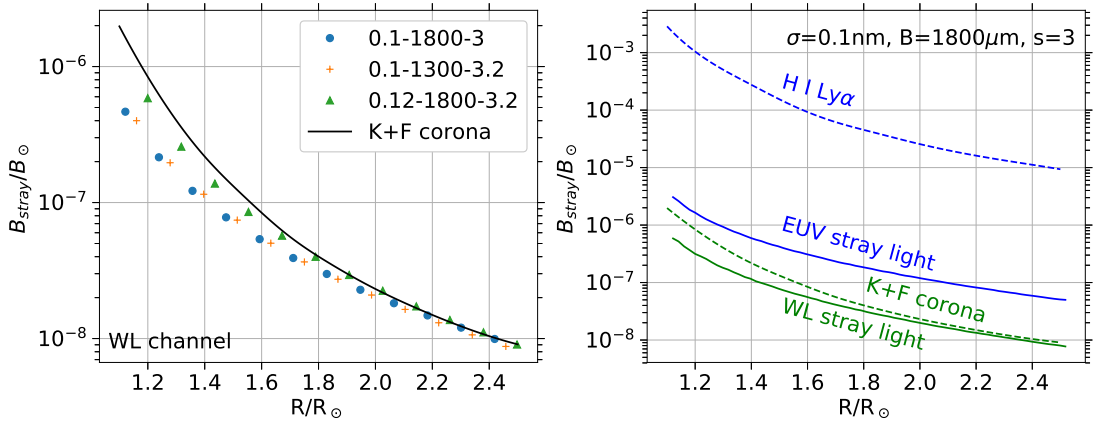}
\end{center}
\caption[Stray light fitting.]
{\label{fig:slfit} Synthetic stray light by optimized M1 surface parameters. (a) is for WL channels with parameters listed in Table~\ref{tab:slfit}. (b) is for both WL and {Ly\textalpha} channels with parameters noted at the top.}
\end{figure}

The results in Section~\ref{subs:level} suggest that with common surface properties, the stray light at far FoV is much brighter than the K + F corona in WL channel. To fulfill the requirements that stray light should be weaker than coronal signal at full FoV, not only the TIS shall be smaller, but also BRDF shall drop earlier and faster along the scattering angles. It requires that the roughness $\sigma$, roll-off spatial frequency $1/B$ and slope $-s$ are smaller. In this section, we aim to get some parameter combinations that fulfill the stray light requirements. A uniform solar disk is still simulated with \num{1e7} rays, and the K-correlation model is used to simulate M1 scattering. Three combinations of M1 surface parameters that satisfy stray light requirements are obtained through optimization, and they are listed in Table~\ref{tab:slfit}. The corresponding stray light distributions are shown in Fig.~\ref{fig:slfit}.

The assumed low surface roughness (\num{0.1} or \SI{0.12}{\nm}) is near the atom's diameter. However, the RMS roughness within \SI{0.1}{\nm} before coating has been reported in Refs.~\citenum{Bruechner1995lasco,Fineschi1994stray}. Slope -3 is used in Lorentzian model and is not uncommon\cite{Fineschi1994stray,Landini2006score}. The B value in \SIrange[range-units=single, range-phrase = --]{1800}{1300}{\um} corresponds to the roll-off angle of \SIrange[range-phrase = --]{0.022}{0.031}{\degree}, near the lower limit of the effective scattering angles and much smaller than \ang{0.5}; the latter was adopted in Ref.~\citenum{Fineschi1994stray}. Some optically finished surfaces exhibit no apparent breakpoint but follow inverse power laws; they are referred to as \emph{fractal surfaces} (P.~106 in Ref.~\citenum{Stover1995scatter}). In that case, roll-off angle $\lambda/B\approx 0$ and $(B \lvert\vec{x}\rvert / \lambda)^2 \gg 1$, Eq.~(\ref{eq:kbrdf}) can be expressed as
\begin{equation} \label{eq:fbrdf}
 \mathrm{BRDF}(\lvert\vec{x}\rvert) = K\frac{\cos\theta_i \cos\theta_s}{\lvert\vec{x}\rvert^s} \, .
\end{equation}
The small roll-off angles that we expected mean that SCI M1 surface should be approximately fractal to fulfill stray light requirements. For $\sigma=0.1\,\rm nm$, $B=\;$\SI{1800}{\um} and $s=3$ as an example, $K$ in Eq.~(\ref{eq:fbrdf}) is about \num{2.0e-10}.

Although the surface that fits parameter combinations given in Table~\ref{tab:slfit} hasn't been reported (as far as we know), LASCO-C1 is also an internally occulted reflecting coronagraph and meets the stray light requirements\cite{Bruechner1995lasco}, which could be explained by such kind of primary mirror surface.

%%%% M2: results are the same, so be quitted %%%%

%%%%%%%%%%%% Conclusions %%%%%%%%%%%%%
\section{Conclusions and Discussion}
\label{sect:conc}
For surfaces with PSD function having a plateau at low spatial frequencies and following an inverse power law at high spatial frequencies, K-correlation model is a good choice to model the surface scattering. Applying common surface parameters on SCI primary mirror, ratios of signal to stray light in both WL and {Ly\textalpha} channels decrease along FoV, and the stray light in WL channel is around one order of magnitude higher than coronal emission at $2.5\,\Ro$. The stray light contributed by M1 scattering of the solar disk light is mainly from the spatial frequencies of \SIrange[range-units=single, range-phrase = --]{6.7e-4}{2.9e-2}{\per\um}, and more attention shall be paid to reduce the roughness within that range. 

To fulfill the stray light requirements, not only M1 surface roughness shall be reduced, but also the roll-off spatial frequency and log-log slope shall be small. In the perspective of scattering, it means that the total scattering shall be low, and BRDF should drop early and fast along the scattering angles. A certain combination of M1 surface parameters, e.g., RMS roughness $\sigma=0.1\,\rm nm$, roll-off angle $\theta_0=0.022\,^\circ$ and log-log slope $-s=-3$, could yield the stray light levels that fulfill the requirements in both the WL and {Ly\textalpha} channels of the SCI.

When conducting polarization brightness (pB) of the solar corona, much stray light may be filtered out in the WL channel. Considering that solar disk light is mainly unpolarized, and the incident angle is small when the disk light is scattered by SCI primary mirror, stray light almost keeps unpolarized and can be removed for the polarization measurements. This could more or less mitigate the stringent requirements of stray light in the WL waveband.

This work is helpful to clarify and understand the requirements of the SCI primary mirror for the sake of stray light and set the requirements of its manufacture, polishing, and coating. The K-correlation model has been incorporated in the Zemax OpticStudio, and is worth considering when looking for a scattering model.

\appendix    %>>>> this command starts appendixes

\section{Definitions of scattering quantities}
\label{sect:defi}
To simplify the problems, all the definitions are based on the 1-D surface. For more detailed definitions, Ref.~\citenum{Stover1995scatter} is recommended. The surface profile is $z=z(x)$, with the surface length of $L_s$. The surface profile is usually described by the following functions:
\begin{itemize}
 \item The surface autocovariance function:
  \begin{equation} \label{eq:acv}
   G(\tau) = \lim_{L_s\to \infty}\frac{1}{L_s}\int_{-L_s/2}^{L_s/2}\left[z(x)-\bar{z}\right]\left[z(x+\tau)-\bar{z}\right]\ud x \, .
  \end{equation}
 \item The autocorrelation function:
  \begin{equation} \label{eq:autof}
   C(\tau) = \lim_{L_s\to \infty}\frac{1}{L_s}\int_{-L_s/2}^{L_s/2}z(x)z(x+\tau)\ud x = G(\tau)+\bar{z}^2 \, .
  \end{equation}
 \item The power spectral density (PSD) function:
  \begin{equation} \label{eq:psdf}
   \mathrm{PSD} = S_1(f_x) = \lim_{L_s\to \infty}\frac{1}{L_s}\lvert Z(f_x,L_s)\rvert^2=\lim_{L_s \to \infty} \frac{1}{L_s} \lvert\int_{-L_s/2}^{L_s/2}z(x)e^{-j2\pi f_x x}\ud x\rvert^2 \, .
  \end{equation}
\end{itemize}
In Eqs.~(\ref{eq:acv}-\ref{eq:psdf}), $\bar{z}$ is the average of $z(x)$. The surface PSD function is the Fourier transform pair of the autocorrelation function:
\begin{equation} \label{eq:ftpsd}
 S_1(f_x) = \int_{-\infty}^{\infty} C(\tau) e^{-j2\pi f_x \tau} \ud\tau = \int_{-\infty}^{\infty} G(\tau) e^{-j2\pi f_x \tau} \ud\tau + \bar{z}^2\delta(f_x) \, .
\end{equation}

The root mean square (RMS) surface roughness $\sigma_{\rm total}$ and autocorrelation length $l_c$ are respectively defined by
\begin{equation} \label{eq:sigmat}
 \sigma_\mathrm{total} = \left(\lim_{L_s\to \infty}\frac{1}{L_s}\int_{-L_s/2}^{L_s/2}\left[z(x)-\bar{z}\right]^2\ud x\right)^{1/2} = \sqrt{G(0)} \, ,
\end{equation}
and
\begin{equation} \label{eq:lc}
 G(l_c) = \frac{G(0)}{e} = \frac{\sigma_\mathrm{total}^2}{e} \, .
\end{equation}

For a grating surface $z(x) = a\sin(2\pi f_x x)$, $G(\tau)$ is
\begin{equation} \label{eq:acvg}
 G(\tau) = \frac{a^2}{2}\cos(2\pi f_x\tau) \, ,
\end{equation}
and
\begin{equation} \label{eq:slc}
 \sigma_\mathrm{total} = \frac{a}{\sqrt{2}} \, , \quad l_c=\frac{1}{2\pi f_x}\arccos(\frac{1}{e}) \, .
\end{equation}
Equation~(\ref{eq:slc}) indicates that for a grating surface, $\sigma_{\rm total}$ is only related to the surface amplitude, and $l_c$ only to the spatial frequency. By Fourier's theorem, any function can be synthesized by a series of harmonic functions. Thus the example of the grating surface is not only simple, but also important to understand other kinds of surfaces.
Additionally, $\sigma_{\rm total}$ can be derived from the surface PSD (or band-limited RMS roughness for a limited integral range $f_{min}-f_{max}$):
\begin{equation} \label{eq:sigmat2}
 \sigma_\mathrm{total}^2 = \int_{-\infty}^{\infty} S_1(f_x) \ud f_x = 2\int_{0}^{\infty}S_1(f_x)\ud f_x \, .
\end{equation}

The BRDF and TIS are defined by
\begin{equation} \label{eq:brdfd}
 \mathrm{BRDF} \equiv \frac{\mathrm{differential~radiance}}{\mathrm{differential~irradiance}} \simeq \frac{\mathrm{d}P_s/\mathrm{d}\Omega_s}{P_i \cos\theta_s} \, ,
\end{equation}
and
\begin{equation} \label{eq:tisd}
 \mathrm{TIS} \equiv \frac{P_s}{RP_i} \, ,
\end{equation}
respectively, where $P$ is light power and $\Omega$ is solid angle. Assuming the surface reflectivity $R=1$, Eq.~(\ref{eq:tis}) is obtained.

\acknowledgments % equivalent to \section*{ACKNOWLEDGMENTS}       
 
We thank Lingping He and Quanfeng Guo for useful discussions. Jianchao Xue gratefully acknowledges the support of University of Florence for his visit from March 2018 to February 2019. This work is supported by 
Natural National Science Foundation of China (NSFC) (11427803, U1731241); CAS Strategic Pioneer Program on Space Science (XDA15052200, XDA15320103, XDA15320301).
% References
\bibliography{StrayL_spie} % bibliography data in report.bib
\bibliographystyle{spiebib} % makes bibtex use spiebib.bst

\end{document}